\documentclass{aa}  
\usepackage{natbib}
\usepackage{graphicx}
\usepackage{txfonts}
\usepackage{hyperref}
\begin{document} 
  \title{New transient Galactic bulge intermediate polar candidate XMMU\,J175035.2-293557}
  \author{    F.~Hofmann\inst{1}
     \and     G.~Ponti\inst{1}
     \and     F.~Haberl\inst{1}
     \and     M.~Clavel\inst{2}
       }

\titlerunning{The intermediate polar candidate XMMU\,J175035.2-293557}
\authorrunning{Hofmann et al.}

\institute{Max-Planck-Institut f\"ur extraterrestrische Physik, Giessenbachstra{\ss}e, 85748 Garching, Germany
           \and Univ. Grenoble Alpes, CNRS, Institut de Planétologie et d'Astrophysique de Grenoble (IPAG), 38000, Grenoble, France
          }

  \date{Received ..., accepted ...}
  
  \abstract{For the past decades a rare subclass of cataclysmic variables (CV), with magnetized white dwarfs (WD) as accretors has been studied and called intermediate polars (IP). They have been discussed as the main contributor to the diffuse X-ray emission due to unresolved point sources close to the Galactic center (GC) and in the Galactic bulge (GB).}
  {In an ongoing X-ray survey (0.5-10\,keV energy band) of $\mathrm{3\degr \times 3\degr}$ around the GC with the \emph{XMM-Newton} observatory we conducted a systematic search for transient X-ray sources.}
  {Promising systems were analyzed for spectral, timing, and multi-wavelength properties to constrain their nature.}
  {We discovered a new highly variable (factor $\mathrm{\gtrsim{20}}$) X-ray source about $\mathrm{1.25\degr}$ south of the GC. We found evidence making the newly discovered system a candidate IP. The X-ray light curve shows a period of $\mathrm{511\pm10~s}$ which can be interpreted as the spin period of the WD. The X-ray spectrum is well fit by a bremsstrahlung model with a temperature of $\mathrm{13.9\pm2.5~keV}$, suggesting a WD mass of $\mathrm{0.4-0.5~M_\odot}$. Among many candidates we could not identify a blue optical counterpart as would be expected for IPs.}
  {The high X-ray absorption and absence of a clear optical counterpart suggest that the source is most likely located in the GB. This would make the system a transient IP (GK Per class) with especially high peak X-ray luminosity and therefore a very faint X-ray transient (VFXT).}


\maketitle

\section{Introduction}

Intermediate polars (IP) are a subclass of magnetic cataclysmic variables (CV) where a white dwarf (WD) is accreting mass from a late-type donor star \citep[see e.g.][]{2001PASP..113..764D,2017PASP..129f2001M}. The accretion disk around the WD is interrupted by the magnetic field and matter is accreted along the field lines onto the magnetic poles of the WD. X-ray emission is created in accretion curtains onto the magnetic poles and the accretion disk of the WD \citep[e.g.][]{1994PASP..106..209P,2007IAUS..243..325H,2017A&A...600A.105B}.
IPs usually show light curve modulations (X-ray and optical) by the spin of the WD and the orbital period around its donor star \citep[see e.g.][]{1985MNRAS.212..917W,1988MmSAI..59..117O,1990SSRv...54..195C,1995A&A...298..165K}.
There are about 50 confirmed IPs and more than 100 candidates known today (see IP catalogue maintained by Koji Mukai\footnote{\url{https://asd.gsfc.nasa.gov/Koji.Mukai/iphome/}}). 

Recently IPs have been discussed as the origin of the hard ($\mathrm{\sim2-40~keV}$ range) X-ray emission close to the Galactic center (GC) and in the Galactic bulge (GB) region \citep[e.g.][]{2007A&A...463..957K,2009Natur.458.1142R,2015Natur.520..646P,2016ApJ...826..160H}. \citet{2014MNRAS.442.2580P} estimated the space density of IPs from a \emph{Swift}-BAT survey (14-195\,keV), where due to their high temperature X-ray spectrum, they are brighter than in the 0.5-10\,keV band.
\citet{2004ApJ...613.1179M}, \citet{2005ApJ...634L..53L}, and \citet{2006ApJ...640L.167R} derived from a deep \emph{Chandra} survey that most fainter, hard GC X-ray sources should be IPs.
Investigating the population of IPs in the GB/GC area is therefore very important for understanding their contribution to the diffuse X-ray emission \citep[][]{2012MNRAS.427.1633H}. In addition IPs trace the CV population \citep[see][]{1990SSRv...54..195C} which is important to understand the GB stellar composition \citep[e.g.][]{2014ApJ...790..164C,2015ApJ...810....8C}.

Because IPs are relatively faint in X-rays and optical and due to high extinction towards the GC, deep and high spatial resolution observations are needed.
We used the extension of the deep \emph{XMM-Newton} GC X-ray survey \citep[about 0.5-10\,keV energy band,][]{2015MNRAS.453..172P} to the GB (Ponti et al., in prep.) to search for transients and highly variable sources (overall $\mathrm{\sim3\degr \times 3\degr}$).
We detected a new X-ray source $\mathrm{1.25\degr}$ south of Sgr\,A$^\star$ (in Galactic coordinates). The analysis of the X-ray spectrum, search for periodic modulations and lack optical counterpart identification make the source an IP candidate most likely located in the GB, which are rarely observed \citep[see][]{2009ApJ...699.1053H,2013ApJ...769..120B,2014MNRAS.440..365T,2017MNRAS.466..129J}. A larger sample of clearly identified IPs in the GB will be needed for direct population constraints. As a luminous, transient IP, the source would be the second member of the GK Per class \citep[see recent description by][]{2016MNRAS.459..779Y}, and might have a subgiant donor star.

\section{X-ray data}

The analysis in this work is based on a 25\,ks \emph{XMM-Newton} observation (ObsID: 0801681401, start date 2017-10-07, PI: Ponti). We used data from the \emph{XMM-Newton} European Photon Imaging Camera (EPIC) PN \citep[][]{2001A&A...365L..18S}, and MOS CCDs \citep[][]{2001A&A...365L..27T}. The spectral fitting was performed using \texttt{XSPEC} \citep[version 12.9.1n,][]{1996ASPC..101...17A} and the Bayesian fitting package BXA \citep[][]{2014A&A...564A.125B}. 
The data reduction was performed using the \emph{XMM-Newton} Science Analysis System (SAS) version 16.1.0. Long-term source variability information was obtained together with previous 15\,ks and 2\,ks \emph{Chandra} \citep[description see][]{2003SPIE.4851...28G} observations (ObsIDs: 7167, start date 2006-10-30, PI: Grindlay; 8753, start date 2008-05-13, PI: Jonker), which were part of the \emph{Chandra} Galactic bulge survey \citep[][]{2011ApJS..194...18J}.
These observations were reprocessed using the \emph{Chandra} Interactive Analysis of Observations software package \citep[CIAO;][]{2006SPIE.6270E..1VF} version 4.5 and the \emph{Chandra} Calibration Database \citep[CalDB;][]{2007ChNew..14...33G} version 4.5.9.
Uncertainties are quoted on the $\mathrm{1\sigma}$ level unless stated otherwise. 

\section{Results}

The source position is R.A.: 17$\rm{^h}$50$\rm{^m}$35\fs2 Dec.: -29\degr35\arcmin57\farcs2 with an uncertainty of 1.2\arcsec\ (systematic, see \emph{XMM-Newton} technical note: XMM-SOC-CAL-TN-0018) plus 0.5\arcsec\ (statistical), leading to 1.3\arcsec\ total $\mathrm{1\sigma}$ uncertainty. 
A cross correlation of detected sources in the \emph{XMM-Newton} observation with the \emph{Chandra} Galactic Bulge survey catalogue of X-ray and optical sources \citep[][]{2014ApJS..210...18J,2016MNRAS.458.4530W} showed an average offset of less than 1\arcsec\ in both R.A. and Dec. direction.

\begin{figure}[ht]
    \centering
    \includegraphics[width=0.49\textwidth,angle=0,trim=1cm 14cm 0.5cm 6cm,clip=true]{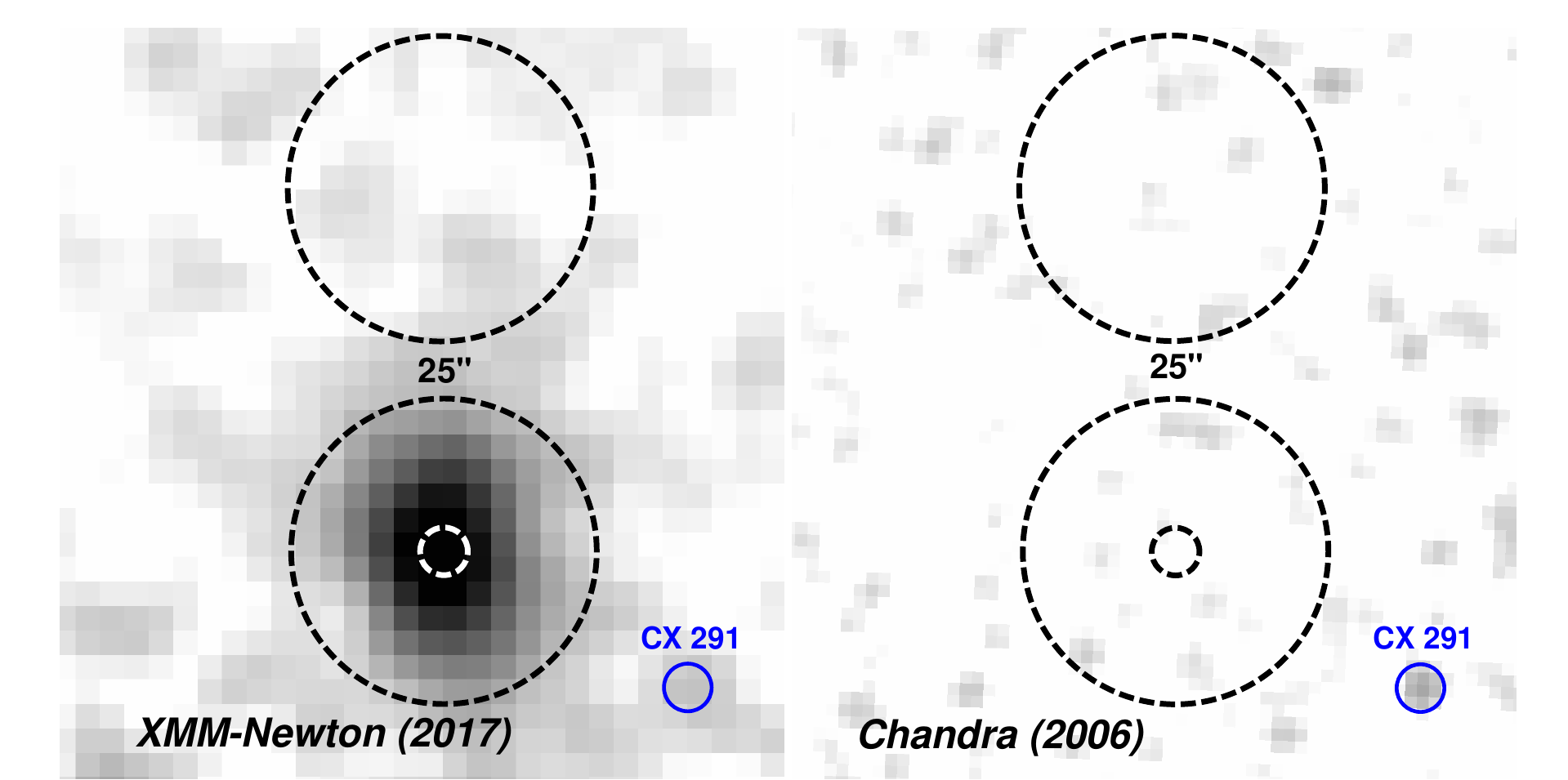}
    \includegraphics[width=0.49\textwidth,angle=0,trim=1cm 9.5cm 0.5cm 11.5cm,clip=true]{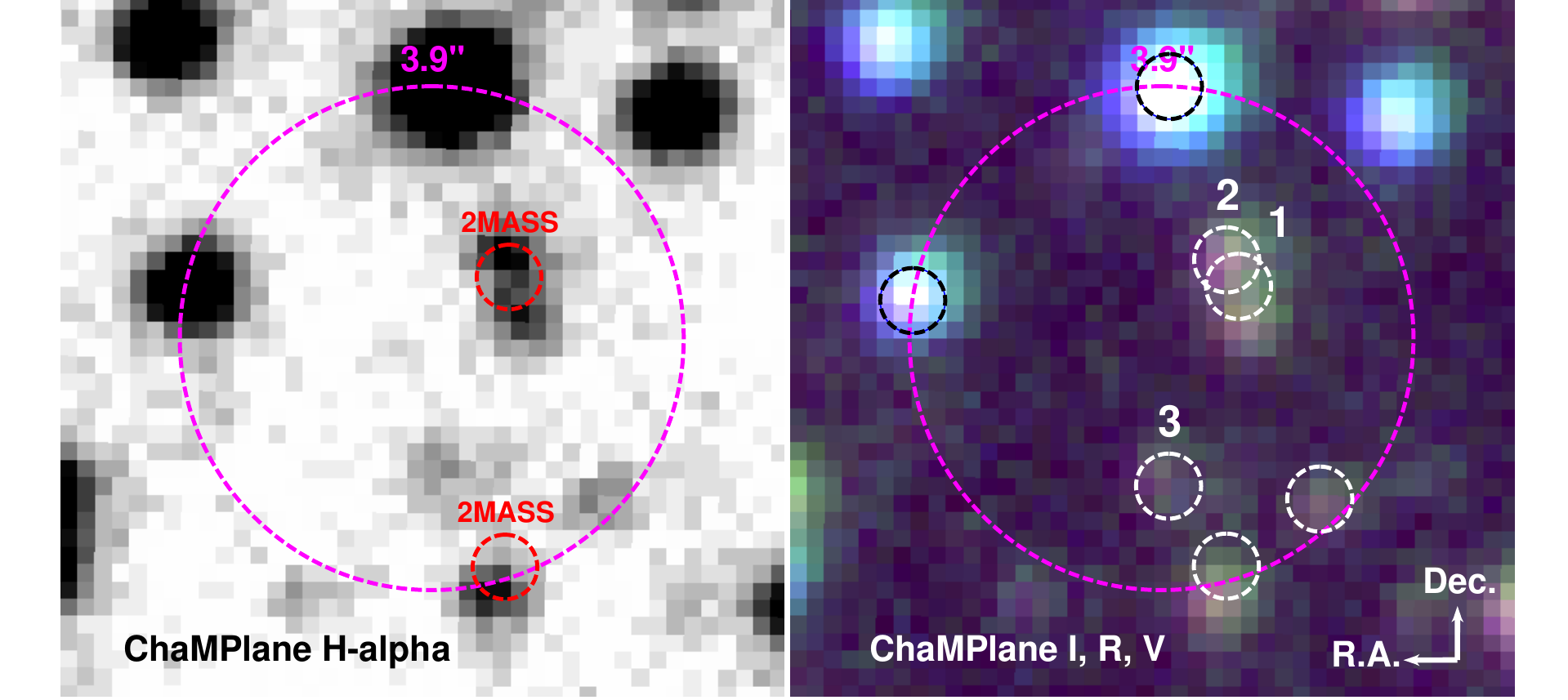}
    \caption{{\bf Top left:} \emph{XMM-Newton} EPIC 1-2\,keV flux image (from 2017-10-07, $\mathrm{4\times4}$\arcsec\ pixel size, smoothed by a 2 pixel Gaussian). Circular extraction regions (25\arcsec\ radius) of source and background counts (black dashed), $\mathrm{3\sigma}$ positional uncertainty (white dashed). {\bf Top right:} \emph{Chandra} ACIS-I 1-2\,keV flux image (from 2006-10-30, $\mathrm{2\times2}$\arcsec\ pixel size, smoothed by a 2 pixel Gaussian). The blue circle indicates a source from the \citet{2014ApJS..210...18J} \emph{Chandra} catalogue. {\bf Bottom left:} Zoomed ChaMPlane \citep[][]{2003AN....324...57G} $\mathrm{H_\alpha}$ image (red circles: 2MASS sources, magenta circle: $\mathrm{3\sigma}$ X-ray positional uncertainty). {\bf Bottom right:} ChaMPlane colour band image (red: I, green: R, blue: V). Small circles are correlated sources from \citet{2016MNRAS.458.4530W} with the three closest numbered.}
    \label{fig:findingChart}
\end{figure}

\subsection{X-ray spectrum and variability}

\begin{figure}[ht]
    \centering
    \includegraphics[width=0.5\textwidth,angle=0,trim=0cm 0cm 0cm 0cm,clip=true]{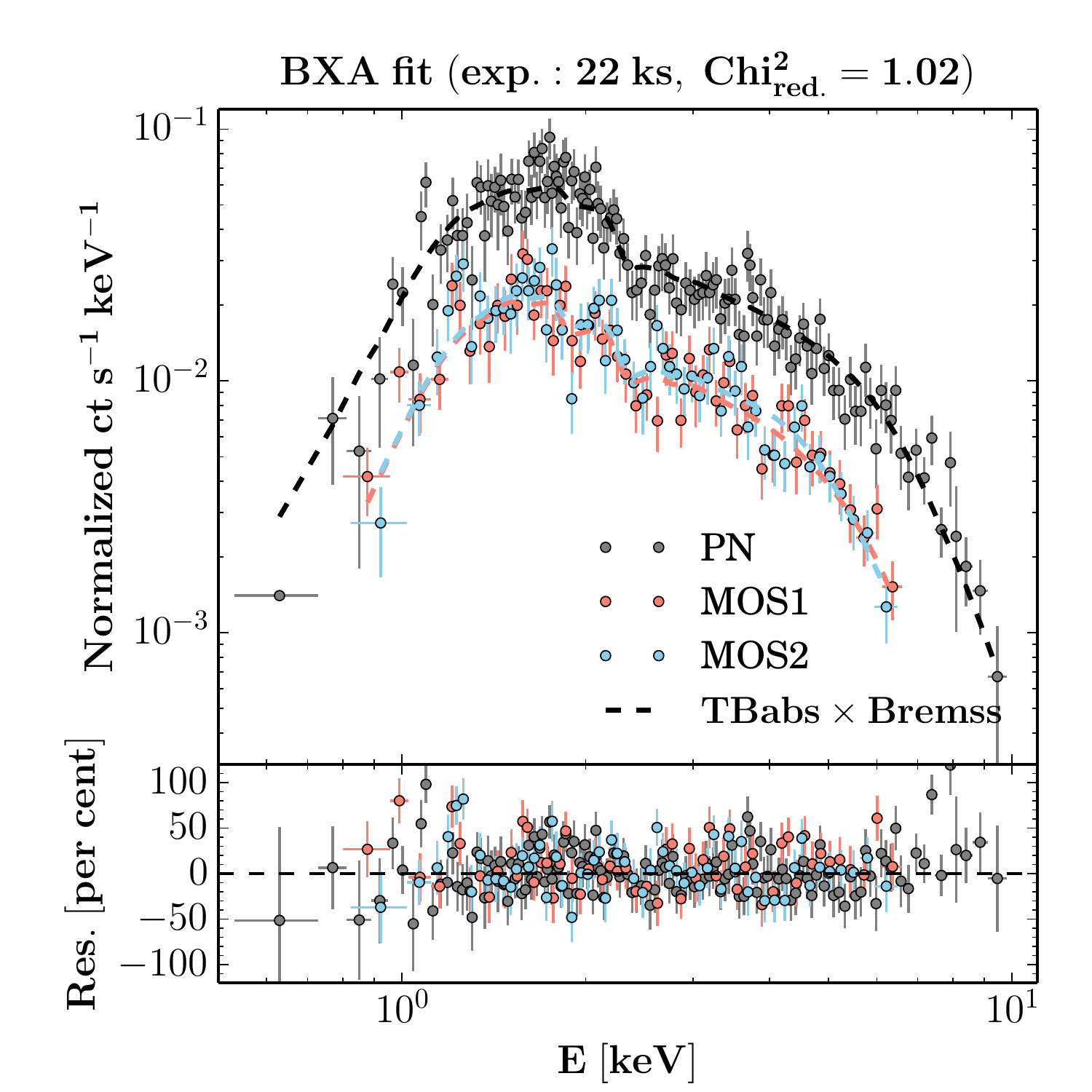}
    \caption{\emph{XMM-Newton} EPIC spectrum (0.5-10\,keV) for PN, MOS1, and MOS2 separately with residuals of a \texttt{TBabs}$\times$\texttt{bremss} model fitted using BXA in \texttt{pyXSPEC}.}
    \label{fig:spectrum}
\end{figure}

\begin{table}[h]
\caption[]{\emph{XMM-Newton} spectral model parameters (0.5-10\,keV range).}
\begin{center}
\begin{tabular}{llllll}
\hline\hline\noalign{\smallskip}
  \multicolumn{1}{l}{} &
  \multicolumn{1}{l}{\texttt{pow}\tablefootmark{a}} &
  \multicolumn{1}{l}{\texttt{diskbb}\tablefootmark{a}} &
  \multicolumn{1}{l}{\texttt{bremss}\tablefootmark{a}} &
  \multicolumn{1}{l}{\texttt{apec}\tablefootmark{a,}\tablefootmark{c}} \\
\noalign{\smallskip}\hline\noalign{\smallskip}
  $\mathrm{N_H\tablefootmark{b}}$ & $3.6\pm0.2$ & $1.0\pm0.1$ & $1.2\pm0.1$ & $1.3\pm0.1$ \\
  $\mathrm{kT\tablefootmark{b}}$ & - & $2.2\pm0.1$ & $13.9\pm2.5$ & $15.5\pm2.5$ \\
  $\mathrm{\Gamma\tablefootmark{b}}$ & $2.8$ & - & - & - \\
  $\mathrm{F_X\tablefootmark{b}}$ & $1.6$ & $1.7$ & $1.7$ & $1.7$ \\
  $\mathrm{\chi^2/dof\tablefootmark{b}}$ & 619/260 & 302/260 & 265/260 & 273/258 \\
\noalign{\smallskip}\hline
\end{tabular}
\tablefoot{
\tablefoottext{a}{Implemented with BXA fitting in \texttt{pyXSPEC}.}
\tablefoottext{b}{Hydrogen column density $\mathrm{N_H~[10^{22}~cm^{-2}]}$, temperature $\mathrm{kT~[keV]}$, power law photon index $\mathrm{\Gamma}$, X-ray flux $\mathrm{F_X~(0.5-10~keV)~10^{-12}~erg~s^{-1}~cm^{-2}}$ (uncertainties $\mathrm{\sim3\times10^{-14}~erg~s^{-1}~cm^{-2}}$), and goodness of fit ($\mathrm{\chi^2}$) with degrees of freedom (dof) in the fit.}
\tablefoottext{c}{metal abundance fixed to solar.}
}
\end{center}
\label{tab:fit}
\end{table}

We extracted the source and background spectra (0.5-10\,keV) from two 25\arcsec\ radius circles shown in Fig. \ref{fig:findingChart}, using standard event file filtering and grouped the spectrum to contain at least 22\,counts per bin.
The background region was chosen to contain the same level of enhanced X-ray emission as the source region \citep[caused by extended recombining plasma south of the GC, see][]{2013ApJ...773...20N}.
The \emph{XMM-Newton} EPIC spectrum contains $\mathrm{\sim6\times10^3}$ net counts (0.5-10\,keV range).
Table \ref{tab:fit} shows the best fit parameters and goodness of fit ($\mathrm{\chi^2}$) for several absorbed single-component spectral models.
The fitted parameters were obtained using the BXA fitting software with wide and flat priors. For every tested parameter combination the model flux was calculated. The lower limit, best fit value, and upper limit are the 15, 50, and 85 percentiles of the parameter distributions (transformed into symmetric uncertainties).
The analysis shows that bremsstrahlung (\texttt{bremss}) is the best fitting model compared to a disk-blackbody (\texttt{diskbb}, $\mathrm{\Delta\chi^2=37}$) or powerlaw (\texttt{pow}, $\mathrm{\Delta\chi^2=354}$). A collisionally ionized plasma (\texttt{apec}) model with solar abundance would also fit the data well (see Table \ref{tab:fit}).
There is no evidence for an $\mathrm{Fe}$-line complex in the $\mathrm{6-7~keV}$ range. We obtained $\mathrm{3\sigma}$ upper limits on the equivalent width (EW) of emission lines at energies: $\mathrm{EW(6.4~keV) \lesssim 400~eV}$, $\mathrm{EW(6.7~keV) \lesssim 220~eV}$, and $\mathrm{EW(7.0~keV) \lesssim 450~eV}$ \citep[main peaks of the $\mathrm{Fe}$-line complex, see e.g.][]{2004MNRAS.352.1037H} by adding additional narrow Gaussian lines to the \texttt{bremss} model.
The measured foreground column density of neutral hydrogen was fitted as $\mathrm{N_H \approx (1.2\pm0.1)\times10^{22}~cm^{-2}}$ using the \texttt{TBabs} model with cross-sections and abundances from \citet{2000ApJ...542..914W}.

Using the best fit $\mathrm{\texttt{TBabs} \times \texttt{bremss}}$ model the \emph{XMM-Newton} count rate of $\mathrm{\sim0.3~ct~s^{-1}}$ translates into a flux of $\mathrm{1.7\times10^{-12}~erg~s^{-1}~cm^{-2}}$ (here 0.5-7.0\,keV, for comparison with \emph{Chandra}).
The $\mathrm{3\sigma}$ upper limit from previous \emph{Chandra} observations of $\mathrm{\sim0.005~ct~s^{-1}}$ translates to $\mathrm{8.6\times10^{-14}~erg~s^{-1}~cm^{-2}}$ (0.5-7.0\,keV). This means the source flux varies by a factor of $\mathrm{\gtrsim{20}}$ on long time scales ($\mathrm{\sim10~yr}$ between upper limit and detection). Within the \emph{XMM-Newton} EPIC PN observation the flux varies significantly ($\mathrm{\sim5\sigma}$) by up to a factor of $\mathrm{\sim4}$ on time scales of minutes (77\,s binning, see Fig.\ref{fig:psd}).
The source is also covered as part of the Swift Galactic bulge survey \citep[][]{2017HEAD...1640001S} with the \emph{Neil Gehrels Swift Observatory} X-ray telescope \citep[mission description, see][]{2005SSRv..120..165B} in 2017 and 2018, but the 60-120\,s exposures did not allow to significantly constrain the long term light curve. Recent Swift ToO observations (ObsIDs: 00010652001/2, 2018-04-08/10) and a renewed coverage within the \emph{XMM-Newton} GC/GB survey (ObsID: 0801682101, 2018-03-18) provide an upper limit of $\mathrm{F_X\lesssim7.2\times10^{-14}~erg~s^{-1}~cm^{-2}}$ (a factor of $\mathrm{\gtrsim{20}}$ fainter again).

\subsection{X-ray periodicity}

\begin{figure*}[ht]
    \centering
    \includegraphics[width=\textwidth,angle=0,trim=0cm 0cm 0cm 0cm,clip=true]{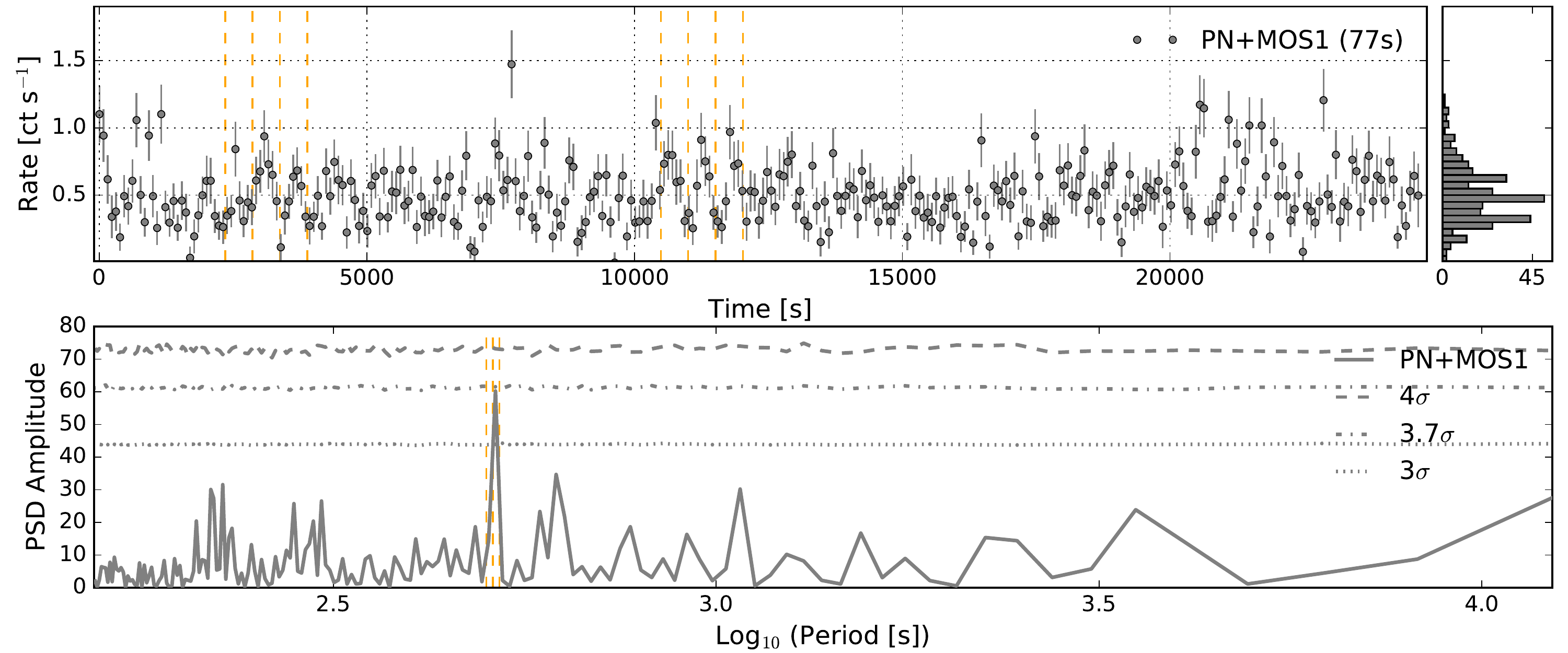}
    \caption{{\bf Top:} Background subtracted, vignetting corrected X-ray ($\mathrm{0.5-10~keV}$) light curve of \emph{XMM-Newton} EPIC PN+MOS1, binned to 77\,s. Dashed orange lines mark intervals where the 511\,s pulsations are visible by eye, the histogram shows the distribution of rates. {\bf Bottom:} PSD amplitude [$\mathrm{Rate^2~Hz^{-1}}$]. Dashed lines show the confidence contours of simulated light curves without periodicity, assuming white noise variability. The orange dashed lines indicate the $\mathrm{511\pm10~s}$ period.}
    \label{fig:psd}
\end{figure*}

We extracted X-ray light curves for source and background from the same regions as the spectra for 24.7\,ks of observation with the PN, MOS1, and MOS2 instruments. We used standard event filtering, barycentric photon arrival time corrections, and the same good time intervals (GTI) to obtain background subtracted, vignetting corrected $\mathrm{0.5-10~keV}$ light curves from all instruments.

We performed a Lomb-Scargle analysis \citep[][]{1982ApJ...263..835S} with the \texttt{periodogram} tool provided by the NASA exoplanet archive\footnote{\url{https://exoplanetarchive.ipac.caltech.edu/docs/tools.html}} and found a peak at $\mathrm{\sim511~s}$ in the unbinned PN data (confirmed by the MOS1 data at lower significance because of its smaller effective area). We discard MOS2 in the following analysis because of higher and less stable background \footnote{MOS2 showed a significantly more noisy background PSD (corrupting the source periodogram), compared to PN and MOS1.}. The period was found independent of the light curve binning. 

For a more detailed significance analysis, 77\,s binning was chosen as a compromise between time resolution and measurement uncertainty in each bin. Fig. \ref{fig:psd} shows the light curve of PN+MOS1 and the power spectral density (PSD) calculated for 161 frequencies (total power: $\mathrm{1213~ct^2~s^{-1}}$) with the Fast Fourier Transform (FFT) \texttt{periodogram} function of the \texttt{scipy} Python package (version 0.17.1).
To estimate the significance of the 511\,s peak we simulated $\mathrm{10^6}$ light curves with root mean square (RMS) amplitude normalized to the measured signal. White noise simulations best reproduced the observed PSD, leading to the flat significance contours seen in Fig. \ref{fig:psd}. The most significant peak is located at $\mathrm{511\pm10~s}$ ($\mathrm{\sim3.7\sigma}$), consistent with the independent Lomb-Scargle analysis (see above).
In addition we performed simulations with a pink to red noise PSD, breaking at a period of 500\,s \citep[typical for CVs, e.g.][]{2014MNRAS.438.1714D}. The results indicate that even in this case the periodicity is significant at $\mathrm{\sim3.9\sigma}$.

\begin{figure}[ht]
    \centering
    \includegraphics[width=0.47\textwidth,angle=0,trim=0cm 0cm 0cm 0cm,clip=true]{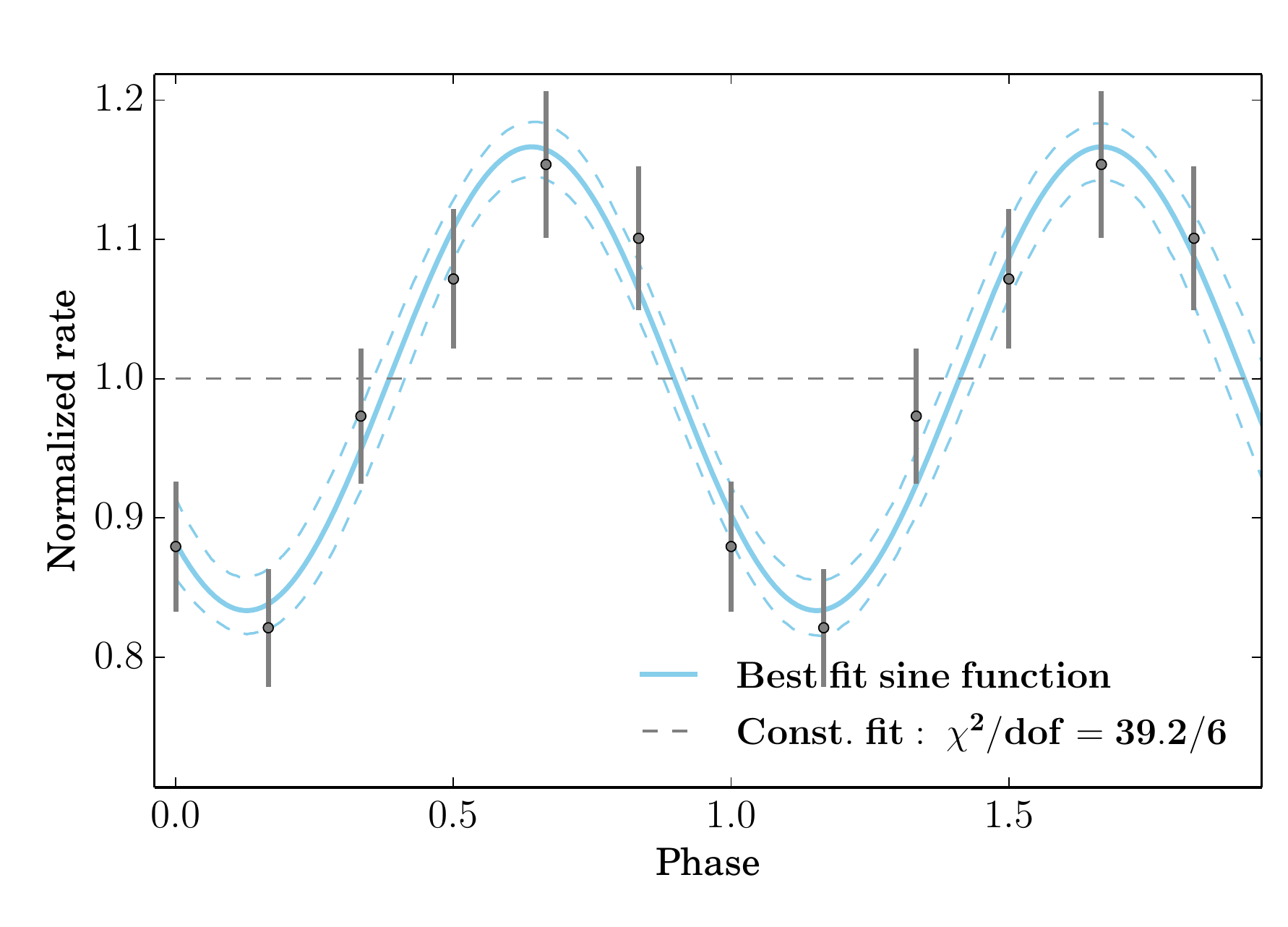}
    \caption{Normalized rate over phase for the \emph{XMM-Newton} EPIC PN light curve (77\,s bins) folded by a 511\,s period. The best fit sine function is displayed in blue (uncertainty from Monte Carlo simulations in dashed blue). The annotations give the goodness of fit to a constant value of 1.0 (dashed grey line).}
    \label{fig:phase}
\end{figure}

Fig. \ref{fig:phase} shows the PN light curve folded by a 511\,s period. The resulting phase diagram is best fit by a sine function with amplitude $\mathrm{0.17\pm0.02}$ (translating into $\mathrm{\sim17~per~cent}$ pulsed fraction). The fit to a constant value results in $\mathrm{\chi^2/dof\approx39.2/6}$ which is further evidence for variability with the folded period.

\subsection{Correlation with optical sources}

The simultaneous \emph{XMM-Newton} optical monitor observation could not be used for counterpart constraints, because the source was outside its field of view.
The closest correlation at $\mathrm\sim1.5\arcsec$ ($\mathrm{\sim1\sigma}$ uncertainty) of the X-ray position is 2MASS\,17503510-2935562 \citep[][]{2003yCat.2246....0C} with $\mathrm{\sim13~(J,H,K)_{mag}}$ (Fig. \ref{fig:findingChart}, bottom left). 
The Yale/San Juan Southern Proper Motion Catalog \citep[][]{2011AJ....142...15G} provides a proper motion of $\mathrm{\sim80~mas/yr}$ for the 2MASS source, which would make it most likely a foreground M star.
The VISTA Variable in the Via Lactea Survey DR2 \citep[][]{2017yCat.2348....0M} provides $\mathrm{\sim 16,15,14,13~(Z,Y,J,H)_{mag}}$ for the position.

The 2MASS position is resolved into two optical sources in the \emph{Chandra} Galactic Bulge survey \citep[][]{2016MNRAS.458.4530W}, and the OGLEII survey \citep[][]{2002AcA....52..217U} - both $\mathrm{\sim 21,17~(V,I)_{mag}}$ (sources 1 and 2, Fig. \ref{fig:findingChart}, bottom). 
The OGLEII light curves (about one observation per day) of possible counterparts show no significant variability, but because of relatively large uncertainties this does not exclude intrinsic variability.

The Gaia DR2 catalogue \citep[][]{2018arXiv180409365G} constrains the distance to source 1 (Fig. \ref{fig:findingChart}, bottom) to $\mathrm{\sim0.8-2.2~kpc}$, and source 2 to $\mathrm{\gtrsim{1.4~kpc}}$, with proper motions of $\mathrm{14~and~11~mas/yr}$ respectively, and both $\mathrm{G_{mag}\sim19}$.

\section{Discussion}

In the \emph{XMM-Newton} GC/GB scan of currently $\mathrm{\sim6\deg^2}$, XMMU\,J175035.2-293557 is a rare intermediate brightness, hard, and strongly variable X-ray source.

\subsection{IP properties from X-rays}

A bremsstrahlung model being the best fit to the X-ray spectrum and the evidence for a 511\,s period are strong indications that the source is an IP.
The best fit temperature of a \texttt{bremss} model is $\mathrm{13.9\pm2.5~keV}$ which would suggest a WD mass of about $\mathrm{0.4-0.5~M_\odot}$ \citep[at the low end of the known mass function, see][]{2003A&A...404..301R,2009A&A...496..121B} with a radius of about $\mathrm{0.01~R_{\odot}}$ and a shock height of the accretion column of $\mathrm{\sim(1-2)\times10^6~cm}$ \citep[][]{2010A&A...520A..25Y}. 
The hard spectrum of the source would fit in the class of IPs with hard X-ray emission \citep[][]{1995A&A...297L..37H}, but the high absorption at lower energies could obscure an additional softer emission. 
The significance of the 500\,s periodicity is above $\mathrm{3\sigma}$ in the soft $\mathrm{0.5-2.0~keV}$ but only $\mathrm{\sim2\sigma}$ in the hard $\mathrm{2.0-10~keV}$ band. This would be expected in IPs, which usually show higher variation amplitude in the softer X-ray band \citep[e.g.][]{2018ApJ...852L...8L}.
The residuals of the bremsstrahlung model (see Fig. \ref{fig:spectrum}) do not show evidence for an $\mathrm{Fe}$-line complex between 6-7\,keV which is often observed for IPs \citep[e.g.][]{2004ApJ...613.1179M,2004A&A...426..253R}. The upper limits on the EW of the main $\mathrm{Fe}$ emission lines are consistent with previously discussed IP candidates \citep[e.g.][]{2009ApJ...699.1053H}.
The X-ray luminosity is in the expected range, but it is among the most distant known IPs with some candidates in the GB \citep[see e.g.][]{2009ApJ...699.1053H,2013ApJ...769..120B,2014MNRAS.440..365T,2017MNRAS.466..129J} or identified from periodicity in extragalactic novae \citep[in M\,31 see e.g.][]{2011A&A...531A..22P}.
The expected orbital period in the IP scenario would be $\mathrm{P_{orb}\gtrsim5000~s}$ \citep[$\mathrm{P_{spin}/P_{orb}\lesssim0.1}$,][]{2004ApJ...614..349N}. 
The light curve shows a hint of modulation on time scales of $\mathrm{\sim5000~s}$, but because the observation time was only 25\,ks no significant detection of an orbital period was found.

\subsection{Source location and luminosity}

The measured X-ray absorption is consistent with the expected total column density of the Galaxy \citep[$\mathrm{N_{H}=N_{HI}+2N_{H_2}\approx9\times10^{21}~cm^{-2}}$,][]{2013MNRAS.431..394W} towards the source position. This provides some evidence that the source is located close to the GC \citep[see e.g. dust layer distribution study by][]{2017MNRAS.468.2532J,2018arXiv180200637J}.
If the source was closer (in front of the main absorbers toward the GC) it would need to have a high intrinsic absorption. 

The expected X-ray luminosity range of IPs is about $\mathrm{3\times10^{29}-5\times10^{33}~erg~s^{-1}}$ \citep[][]{2006ApJ...640L.167R} which constrains the distance to the source from its \emph{XMM-Newton} flux and the \emph{Chandra} upper limit to about $\mathrm{0.1-8~kpc}$.
The average luminosity of the source during the \emph{XMM-Newton} EPIC observation is $\mathrm{(1.3\pm0.1)\times10^{34}~erg/s}$ (0.5-7.0\,keV) assuming a $\mathrm{\texttt{TBabs} \times \texttt{bremss}}$ model (see Fig. \ref{fig:spectrum}) and a distance of 8\,kpc \citep[distance to the GC, see e.g.][]{2003ApJ...597L.121E}. The $\mathrm{3\sigma}$ upper limit from previous \emph{Chandra} observations in this case would be $\mathrm{6.6\times10^{32}~erg/s}$ (0.5-7.0\,keV). 

For a local IP a relatively blue optical counterpart would be expected \citep[][]{2009ApJ...699.1053H} but no candidate fitting this criterion was found in the $\mathrm{3\sigma}$ error circle around the source position (see above).
Assuming that the true counterpart is below the sensitivity limit of current catalogues ($\mathrm{V_{mag} \sim 22}$), with the typical CV magnitude range $\mathrm{M_V \approx 5.5-10.5}$ \citep[e.g.][]{2009ApJ...699.1053H}, and rising absorption from $\mathrm{A_V \approx 1-4}$ \citep[estimated from][]{2011ApJ...737..103S,2014AJ....148...24S} between $\mathrm{1-8~kpc}$, we can constrain its distance to $\mathrm{\gtrsim{1~kpc}}$. Using the hydrogen column density measured in the X-ray spectrum, we estimate $\mathrm{A_V \approx 6}$ \citep[relation by][]{2009MNRAS.400.2050G}.
If the donor star was a subdwarf like in GK Per, one of the faint optical correlations (e.g. number 1, 2, or 3 in Fig. \ref{fig:findingChart}) might be the true counterpart, which would be consistent with location in the GB.
\citet{2009ApJ...699.1053H} found a probable optical counterpart with $\mathrm{21.7~V_{mag}}$ for an IP in the GB, but located in Baade's window where the extinction is only $\mathrm{A_V \approx 1.4}$ at $\mathrm{\gtrsim{3~kpc}}$. 

Together with the $\mathrm{N_H}$, flux, and optical counterpart constraints we used the stellar density and Galactic X-ray source distribution \citep[as shown for the GB IP candidate in][]{2009ApJ...699.1053H} to conclude that the new IP candidate is most likely located in the GB \citep[$\mathrm{8\pm2~kpc}$, see overview of GB structure by][]{2016ARA&A..54..529B}.

\subsection{Alternative interpretations}

Just at the border of the $\mathrm{3\sigma}$ positional error circle is a Be star candidate from the OGLEII survey \citep[][]{2008A&A...478..659S} for which no significant variability was detected. A Be X-ray binary could show periodic pulsations similar to IPs, but the X-ray spectrum would be well fit by a powerlaw with $\mathrm{photon~index\approx1}$ \citep[see e.g. the population of the Small Magellanic Cloud,][]{2004A&A...414..667H,2015MNRAS.452..969C,2016A&A...586A..81H}.

From its X-ray luminosity the newly discovered source is a very faint X-ray transient \citep[VFXT with $\mathrm{L_X\approx10^{34-36}~erg~s^{-1}}$, see e.g.][]{2006MNRAS.366L..31K} and at least a factor of ten flux variation.
The luminosity peaks of the source are around $\mathrm{10^{34}~erg~s^{-1}}$ if it is located at GC distance. The variability within the observation does not indicate an ongoing brightening.
VFXTs are still a poorly explored family of transients that are expected to be mostly X-ray binaries (neutron star or black hole as compact object) with a low accretion rate from a low-mass donor star at $\mathrm{\lesssim10^{-13}~M_\odot/yr}$ \citep[][]{2006MNRAS.366L..31K,2013MNRAS.428.1335M,2015MNRAS.447.3034H}.

We note that the spectrum of XMMU\,J175035.2-293557 is harder than for typical VFXTs \citep[see e.g.][]{2013MNRAS.434.1586A} and periodic pulsations are usually not observed in VFXTs.
Our results show that highly variable IPs could contribute to the faint end of the VFXT population.

\section{Conclusions}

With the results of this work we constrained the nature of the newly discovered X-ray source XMMU\,J175035.2-293557 with the most likely scenario being an intermediate polar in the Galactic bulge. It is among the most luminous candidates, which implies a strong magnetic field of the white dwarf.
The spin period of the white dwarf would be $\mathrm{P_{spin}=511\pm10~s}$, but the orbital period of the system $\mathrm{P_{orb}}$ could not be recovered from the current data. The system was a factor of $\mathrm{\gtrsim{20}}$ fainter in observations $\mathrm{\sim 10~years}$ before and $\mathrm{\sim 6~months}$ after the first detection. This transient nature makes it most likely the second member of the GK Per class.
The X-ray spectra are best fit by a bremsstrahlung model with a temperature of $\mathrm{13.9\pm2.5~keV}$.
No optical counterpart is identified but several candidates are found. The true counterpart might be below the detection threshold of currently available surveys.
With an average flux of $\mathrm{\sim1.7\times10^{-12}~erg~s^{-1}~cm^{-2}}$ this type of source will also be detected in the upcoming eROSITA all-sky survey \citep[][]{2012arXiv1209.3114M} which will allow studying a larger fraction of the Galactic population (on the order of 100 expected).
Following up with high resolution and large effective area X-ray observatories like ATHENA \citep[][]{2013arXiv1306.2307N} the spin and orbital periods of similar IP systems will be significantly detected.

\begin{acknowledgements}

We thank the anonymous referee for very constructive comments that helped to improve the clarity of the paper. 
We thank M. Freyberg and C. Wegg for helpful discussions.
These results are based on observations obtained with \emph{XMM-Newton}, an ESA science mission with instruments and contributions directly funded by ESA Member States and NASA. 
We acknowledge the use of public data from the Swift data archive and thank the Swift team for scheduling the ToO observation; data obtained from the \emph{Chandra} Data Archive and the \emph{Chandra} Source Catalog, and software provided by the \emph{Chandra} X-ray Center (CXC) in the application package CIAO; NASA's Astrophysics Data System; the VizieR catalogue access tool, CDS, Strasbourg, France; SAOImage DS9, developed by Smithsonian Astrophysical Observatory; data and/or software provided by the High Energy Astrophysics Science Archive Research Center (HEASARC), which is a service of the Astrophysics Science Division at NASA/GSFC and the High Energy Astrophysics Division of the Smithsonian Astrophysical Observatory; the SIMBAD database, operated at CDS, Strasbourg, France; NASA's SkyView facility located at NASA Goddard Space Flight Center; the NASA Exoplanet Archive, which is operated by the California Institute of Technology, under contract with the National Aeronautics and Space Administration under the Exoplanet Exploration Program; the NASA/ IPAC Infrared Science Archive, which is operated by the Jet Propulsion Laboratory, California Institute of Technology, under contract with the National Aeronautics and Space Administration; data provided by the Science Data Archive at NOAO. NOAO is operated by the Association of Universities for Research in Astronomy (AURA), Inc. under a cooperative agreement with the National Science Foundation; use of the python packages Matplotlib, scipy, numpy, and pyXSPEC.
F. Hofmann and G. Ponti acknowledge financial support from the BMWi/DLR grants FKZ 50 OR 1715 and 50 OR 1604.

\end{acknowledgements}

\bibliographystyle{aa}
\bibliography{auto}

\end{document}